\documentclass[%
 reprint,
superscriptaddress,
 amsmath,amssymb,
 aps,
pra
]{revtex4-2}

\usepackage{xcolor}
\usepackage{graphicx}
\usepackage{dcolumn}
\usepackage{bm}

\begin{document}

\title{A photonic integrated chip platform for interlayer exciton valley routing}

\author{Kishor K Mandal}
\affiliation{%
Laboratory of Optics of Quantum Materials, Department of Physics, Indian Institute of technology Bombay, Mumbai- 400076, India 
}%
\author{Yashika Gupta}
\affiliation{%
 Laboratory of Optics of Quantum Materials, Department of Physics, Indian Institute of technology Bombay, Mumbai- 400076, India 
}%
\author{Mandar Sohoni}
\affiliation{School of Applied and Engineering Physics, Cornell University, Ithaca, New York 14853, USA}
\author{Achanta Venu Gopal}
\email{achanta@tifr.res.in}
\affiliation{Department of Condensed Matter Physics and Material Science, Tata Institute of Fundamental Research, Homi Bhabha Road, Mumbai 400005, India}
\author{Anshuman Kumar}%
 \email{anshuman.kumar@iitb.ac.in}
\affiliation{%
 Laboratory of Optics of Quantum Materials (LOQM), Department of Physics, Indian Institute of technology Bombay, Mumbai- 400076, India 
}%

\date{\today}

\begin{abstract}
 Interlayer excitons in two dimensional semiconductor heterostructures show suppressed electron-hole overlap resulting in longer radiative lifetimes as compared to intralyer excitons. Such tightly bound interlayer excitons are relevant for important optoelectronic applications including light storage and quantum communication. Their optical accessibility is, however, limited due to their out-of-plane transition dipole moment. In this work, we design a CMOS compatible photonic integrated chip platform for enhanced near field coupling of these interlayer excitons with the whispering gallery modes of a microresonator, exploiting the high confinement of light in a small modal volume and high quality factor of the system. Our platform allows for highly selective emission routing via engineering an asymmetric light transmission which facilitates efficient readout and channeling of the excitonic valley state from such systems.     
\end{abstract}

\maketitle

\begin{figure*}
    \centering
    \includegraphics[width=1\textwidth]{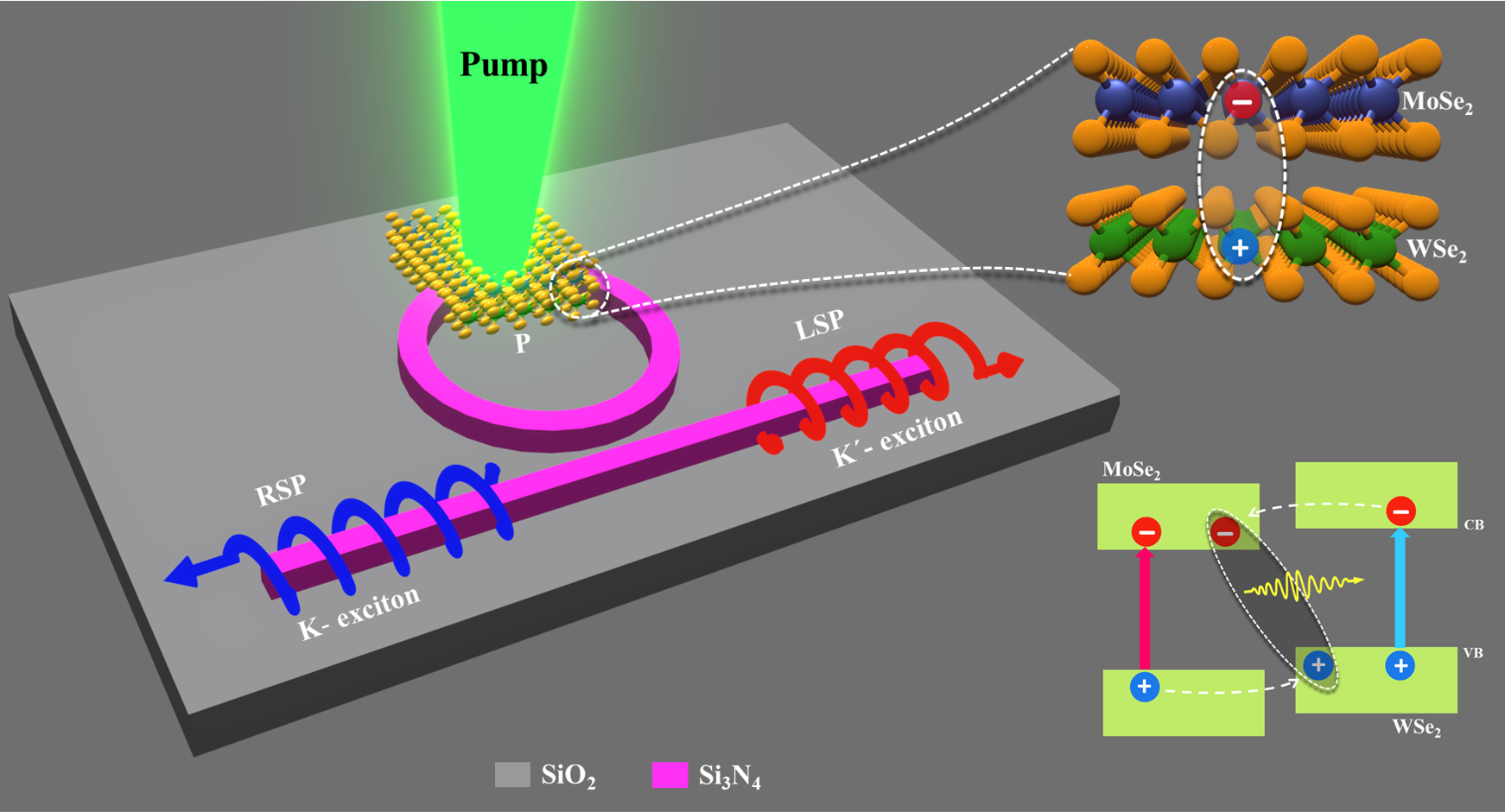}
    \caption{\textbf{Device schematic for interlayer exciton valley routing: }A schematic illustration of valley selective interlayer exciton routing nano-photonic hybrid device showing vertically stacked $\text{MoSe}_{2}/\text{WSe}_{2}$ vdWH integrated on Si$_{3}$N$_{4}$ microring resonator which is coupled to a bus-waveguide. Interlayer ($K$ and $K'$ valley) excitons from vdWH TMDCs get coupled and guided into Si$_{3}$N$_{4}$ microring resonator. These coupled excitons propagate in the opposite direction inside Si$_{3}$N$_{4}$ microring which are evanescently coupled to an adjacent single-mode bus-waveguide. The inset on the top-right corner shows the interlayer exciton in vdWH TMDCs and on the bottom-right shows type-II band alignment in $\text{MoSe}_{2}/\text{WSe}_{2}$ heterostructure facilitating the formation of interlayer excitons.}
    \label{fig:figure1}
\end{figure*}

\emph{Introduction --}Van der Waals heterostructures (vdWH) of atomically thin semiconductors\cite{novoselov20162d} such as transition metal dichalcogenides (TMDCs) have proven to be an effective platform for designing nano and micro-scale optoelectronic and photonic devices for a variety of applications ranging from quantum information processing~\cite{xu2014spin,ReserbatPlantey2021,Baek2020,Wilson2021,Lin2021,Turunen2022} to valleytronics-based devices\cite{Ciarrocchi2022,Schaibley2016,Vitale2018,Langer2018,Mak2018}. 
One promising feature of these heterostructures is their ability to host excitons with large binding energies\cite{Mak2018}. These excitons exhibit various transition dipole moments\cite{Li2020} facilitating a rich diversity of their coupling with various types of optical modes\cite{Chen2019,Krasnok2018,10.1088/1361-6463/ac570e}. Interlayer excitons\cite{Rivera2018}, formed due to the type II band alignment in these heterobilayer structures (shown in the insets of Figure~\ref{fig:figure1}), have recently gained attention due to their longer radiative lifetimes ({$\sim$20--30~ns}) \cite{palummo2015exciton} and smaller scattering rates \cite{surrente2018intervalley}, making such excitons well suited candidates for applications such as quantum information processors and valleytronic devices. 

For valleytronics applications, the peculiar optical selection rules for valley excitons ($K$ or $K'$) of both TMDC monolayers in the vdWH material~\cite{xiao2012coupled, mak2012control, cao2012valley, wang2017plane, zeng2012valley, yu2018brightened} are extremely important. For example, in monolayers, spin-singlet excitons formed in these valleys couple to in-plane circularly polarised photons, resulting from their in-plane transition dipole moment. While their counterpart, spin-triplet excitons, have a weaker dipole transition and couple to linearly $(z)$ polarised photons in the out-of-plane direction. In vdWHs, both spin-singlet and spin-triplet valley excitons can couple with right-handed ($\sigma+$) and left-handed ($\sigma-$) circularly polarized and linearly $(z)$ polarized light\cite{wang2017plane, park2018radiative}. The transition dipole moments of these excitons in vdWHs can be tuned through external electric and magentic fields\cite{Peimyoo2021,Guo2021,Tan2021} or through interlayer translation i.e. based on their atomic registry  in different stacking arrangement  triangular R-type stacking or hexagonal H-type stacking (\cite{rosenberger2020twist, he2021moire}) of vdWH~\cite{yu2018brightened}. As an example, Sohoni et al.~\cite{sohoni2020interlayer} show circularly polarised orthogonal interlayer valley excitonic dipoles in $\rm \text{MoSe}_{2}/\text{WSe}_{2}$ vdWH vertical heterostructure. These interlayer excitons have a tilted quantization axis in both R-type and H-type stacking for certain values of the translation between the two layers. In general, the dipole moments for interlayer valley excitons in the $\pm K$ valleys can be expressed as
\begin{equation}
	\mathbf{D}_{1} = a^{+K}_{+}\mathbf{e}_{+} + a^{+K}_{-}\mathbf{e}_{-} + a^{+K}_{z}\mathbf{e}_{z}
	\end{equation}
	\begin{equation}
	\mathbf{D}_{2} = a^{-K}_{+}\mathbf{e}_{+} + a^{-K}_{-}\mathbf{e}_{-} + a^{-K}_{z}\mathbf{e}_{z}
\end{equation}
{where $\mathbf{D}_{1} = \mathbf{D}_{+K\pm K}$ and $\mathbf{D}_{2} = \mathbf{D}_{-K\mp K}$, depending on the stacking (R/H), $\mathbf{e}_{\pm} = (\hat{x} + i\hat{y})/\sqrt{2}$, and $\mathbf{e}_{z} = \hat{z}$. The coefficients $a^{\pm K}_{\pm,z}$ are dependent on the lateral shift between the monolayers forming the heterostructure. We consider only those cases where the dipoles in the two valleys are orthogonal, that is, $\hat{\mathbf{D}}_{2}^{*}\cdot\hat{\mathbf{D}}_{1} = 0$.} While commensurate TMDC bilayer heterostructures with most of the interlayer translation values, do not exist naturally, they can be found in incommensurate heterostructures with small twist angles (moire superlattices)~\cite{zhang2017interlayer, wang2017interlayer}. These Moiré superlattices also generate an array of such interlayer excitons and thus, have the potential to host interacting quantum emitter arrays which has enormous potential for information processing~\cite{ yu2017moire, seyler2019signatures, jin2019identification, wu2018theory}. However, optically addressing these excitons is a challenge in free space setups due to their out of plane transition dipole moments\cite{Tran2021} and on chip scale due to the weak evanescent overlap in under-optimized structures. This can be overcome by using near field coupling to the optical modes of nanophotonic architectures\cite{Tran2021,Frg2019,Rivera2019,Khelifa2020,Paik2019,Liu2019}. The ‘non-trivial’ tilted circularly polarized transition dipoles result in highly versatile and rich coupling dynamics with such nanophotonic platforms, which can be further tuned by varying the configuration of the bilayer stacking as well as the choice of initial valley excitation\cite{sohoni2020interlayer}.

Separation of valley excitons in TMDC monolayers has so far been implemented via plasmonic nanostructures~\cite{sun2019separation,Wen2020}, dielectric antennae and nanowires\cite{Gong2020,Yao2021} and photonic crystals\cite{Wang2020,PhysRevB.101.245418}. {To the best of our knowledge, there has been no previous work on valley routing for interlayer excitons coupled to nanophotonic architectures.} In this letter, we propose a dielectric microcavity -- ${\rm Si_3N_4}$ - microresonator platform for efficient valley selective separation and routing of interlayer excitons in bilayer $\text{MoSe}_{2}/\text{WSe}_{2}$ heterostructures. Si$_{3}$N$_{4}$ based photonic integrated circuits (PIC) are a fully planar platform which are compatible with complementary metal–oxide–semiconductor (CMOS)-technology\cite{Subramanian2013} and show extremely low loss in the whole visible to infrared spectrum~\cite{kruckel2017optical}. We employ ${ \rm Si_3N_4}$ bus waveguide for routing and ${ \rm Si_3N_4}$ based microring resonator (MRR) as a cavity, which belongs to the class of whispering gallery mode (WGM) type of microcavities, that typically trap photons (spatially and temporally) by re-circulating the resonant mode~\cite{vahala2003optical} and supports high quality factors (high-Q)~\cite{xuan2016high}.

In this work, we explore the coupling between interlayer excitons and ${\rm Si_3N_4}$-microring resonator/bus waveguide system in terms of coupling linewidth, transmission anisotropy, bilayer stacking configuration and polarized excitation for realistic design of interlayer exciton PIC platform. For this purpose, we study the transmission through a bus waveguide coupled to a Si$_{3}$N$_{4}$ microcavity with a vertically stacked $\text{MoSe}_{2}/\text{WSe}_{2}$ vdWH TMDC integrated on top of the microcavity, using  finite difference eigenmode solver (FDE) and three dimensional finite-difference-time-domain (3D-FDTD) simulations via commercially available Ansys Lumerical simulation software. We observe an asymmetric transmission at the bus waveguide end ports with changing helicity of the interlayer exciton dipole, thus opening avenues for a PIC platform for valley state routing and manipulation for interlayer excitons. 

\begin{figure*} [h]
    \centering
    \includegraphics[width=1\textwidth]{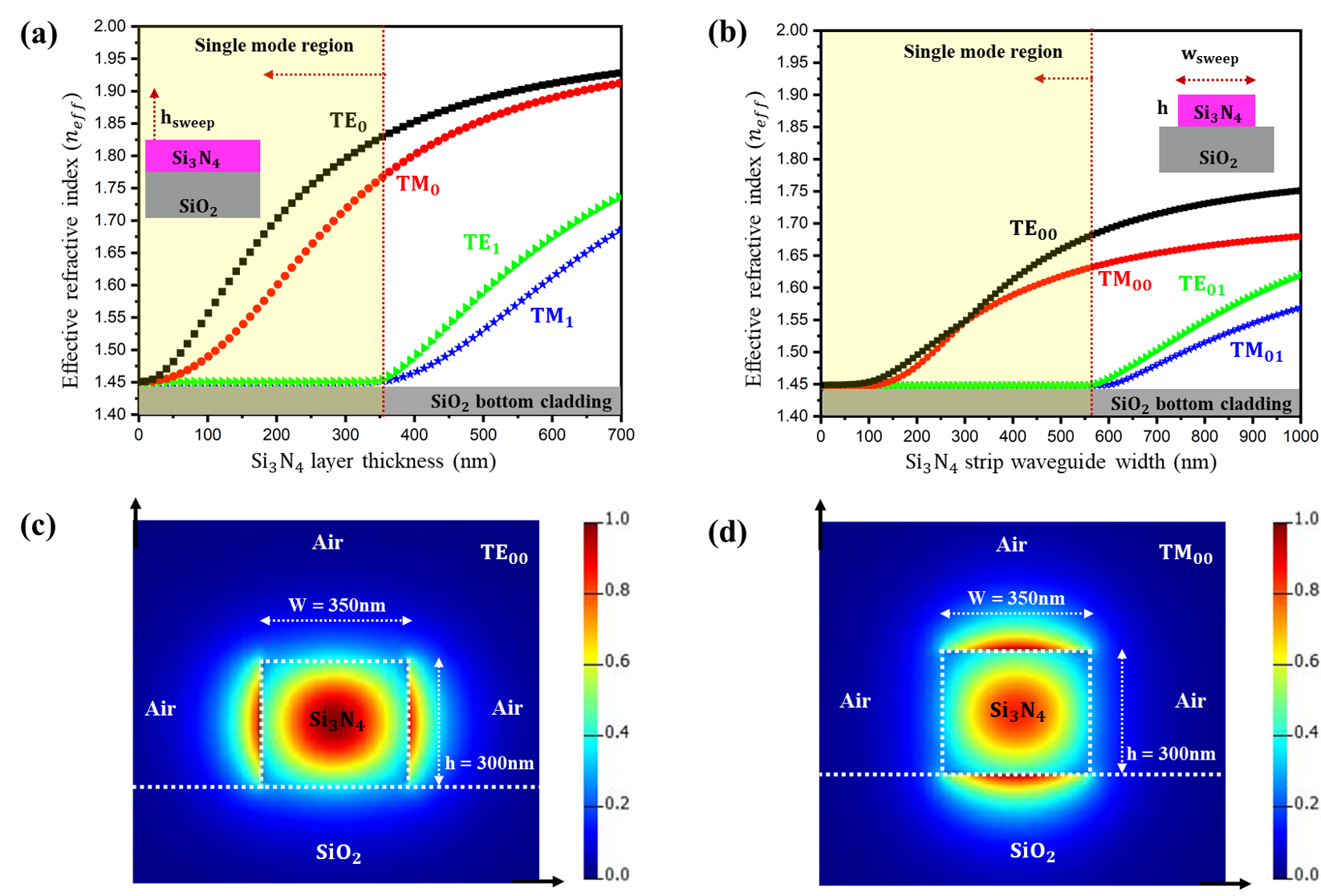}
    \caption{\textbf{PIC optimization:} Mode analysis for obtaining single-mode (demarcated by shaded part) in Si$_{3}$N$_{4}$ strip waveguide using FDE solver at 908.48~nm wavelength. (a) and (b) Show variation of effective refractive index, $ n_{eff}$, with height and width (keeping height constant at 300~nm from (a)), respectively, for modes upto first order. (c) and (d) Show the mode profile for the fundamental quasi-TE$_{00}$ and quasi-TM$_{00}$ polarizations, respectively.}
    \label{fig:figure2}
\end{figure*}

\emph{Designing single mode waveguides for interlayer excitonic PICs --}
We first optimize our PIC platform for single mode operation for the bus waveguide and resonator. These single mode waveguides will help avoid competition with higher-order modes~\cite{han2018simulation}. In our simulation set-up, the geometry consists of a strip-type waveguide with Si$_{3}$N$_{4}$ as the core material on top of a 2-$\mu$m-thick silicon-di-oxide (SiO$_{2}$) layer, acting as the bottom cladding layer to avoid the field  leakage  into the underlying  Si-substrate. The standard Palik model \cite{palik1998handbook} was used for SiO$_{2}$ layer while the optical constant for Si$_{3}$N$_{4}$ was modelled using the Philipp model \cite{philipp1973optical}. We carried out the modal analysis for $1^{st}$ R-type stacking excitonic dipole moment of emission wavelength at 908.48~nm. Firstly, the height $(h)$ of the waveguide was optimized using  a one-dimensional finite difference eigenmode solver (1D-FDE) such that it supports  only a single fundamental mode. Our simulation results indicate that the waveguide with a height up to 360~nm operates in the single mode regime as shown in the shaded region of Figure~\ref{fig:figure2}(a). For this work, we have therefore chosen the height of 300~nm for the waveguide. Secondly, waveguide width $(w)$ optimization was carried out by sweeping the waveguide width up to 1~$\mu$m using two-dimensional 2D-FDE solver, keeping the height constant at 300~nm. For simplicity, the effective refractive index, $\rm n_{eff}$, values till first order modes were measured. Figure~\ref{fig:figure2}(b) shows the effective refractive indices for the fundamental transverse electric quasi-TE$_{00}$, transverse magnetic quasi-TM$_{00}$, and the first order higher modes TE$_{01}$, TM$_{01}$. It can be clearly seen that for widths up to 560~nm, the waveguide exhibits only the fundamental quasi-TE$_{00}$ and  quasi-TM$_{00}$ modes and this width also sets the cut-off for higher order modes (shown by vertical brown-dotted line in Figure~\ref{fig:figure2}(b)). Thus, we choose the single mode operational Si$_{3}$N$_{4}$ strip waveguide with a cross-sectional dimension of 300~nm $\times$ 350~nm ($h \times w$). The cross-sectional mode field profiles of the fundamental quasi-TE$_{00}$ and  quasi-TM$_{00}$ mode supported by the waveguide are depicted in Figure~\ref{fig:figure2}(c) and (d). \\

\emph{Engineered asymmetry for excitonic emission--}
The valley dependent asymmetric transmission of emitted photons from the interlayer exciton recombination was studied using 3D FDTD simulations where, a hybrid integration of stacked monolayers of $\text{MoSe}_{2}$ and $\text{WSe}_{2}$ vdWH TMDCs was kept on top of Si$_{3}$N$_{4}$ microring resonator coupled to a bus-waveguide geometry was considered. The vdWH was placed at a symmetric position, marked as $P$ on the Si$_{3}$N$_{4}$ ring of 30~$\mu$m diameter as shown in Figure~\ref{fig:figure1}. A 30~$\mu$m diameter of the ring was chosen to minimize the bending losses in the propogating cavity-coupled modes and to obtain free spectral range FSR of $\approx 4$~nm    for the resonant mode at wavelength 908.48~nm and n~=~2 refractive index of Si$_{3}$N$_{4}$. The interlayer valley exciton formed in the TMDC stack was modelled as an electric dipole, which also acts as the optical point source for exciting the system.  In the main text, we have presented the detailed analysis of the first spin-singlet R-type stacking configuration of excitonic transition dipole having 908.48~nm emission wavelength. The analysis for remaining dipoles is presented in the supplementary information.

\begin{figure*} [h]
    \centering
    \includegraphics[width=1\textwidth]{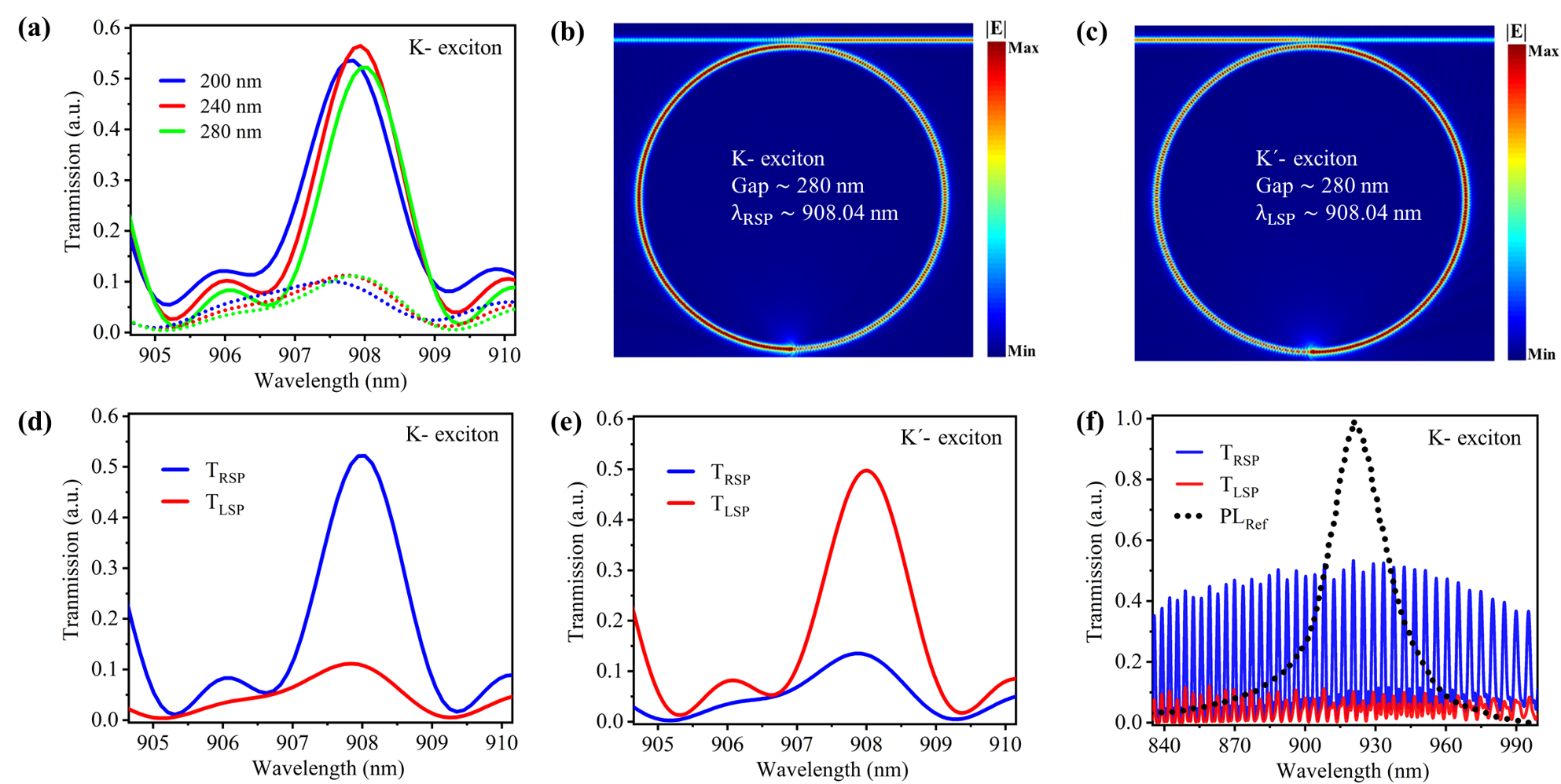}
    \caption{\textbf{Figures of merit for interlayer exciton valley routing: } Study of the asymmetric transmission response of the system for coupled  $K$ and  $K'$ valley interlayer excitons of vdWH TMDCs corresponding to $1^{st}$ R-type singlet exciton emitting at 908.48~nm. (a) Shows fine tuning of the coupling gap to set the transmission peak at the targeted wavelength of 908.48~nm. (b) and (c) Represent the simulated electric field profile, $\rm |E|$ for the propagated K and K' valley excitonic modes, respectively. (d) and (e) Show spectroscopic transmission spectrum captured at LSP (${\rm T_{LSP}}$, red line) and RSP (${\rm T_{RSP}}$, blue line) of the bus-waveguide for K and K' valley excitons, respectively. (f) Demonstrates the broadband spectrum of the cavity resonant peaks (characterized by FSR) overlapped by room temperature free-space photoluminescence (PL) spectrum of the desired interlayer excitons~\cite{khelifa2020coupling}.}
    \label{fig:figure3}
\end{figure*}

The interlayer valley exciton emission is coupled to the Si$_{3}$N$_{4}$ MRR and propagates in both clockwise and counter-clockwhise direction inside the cavity from the originating point $P$ as is shown in Figures~\ref{fig:figure3}(b) and (c). For the $K$ valley exciton emitting at 908.48~nm, the counterclockwise propagating mode is relatively intense in comparison to the clockwise propagating mode as is clearly evident from the electric field ($ \rm |E|$) distribution profile shown in Figure~\ref{fig:figure3}(b), while for the corresponding $K'$ valley exciton, clockwise propogating mode dominates, as shown in Figure~\ref{fig:figure3}(c). These clockwise and counterclockwise travelling modes get coupled out to the bus-waveguide across the gap between the bus and microring and are captured at the right-side port (RSP) and left-side port (LSP) of the bus-waveguide, respectively. For a quantitative analysis of the valley selective transmission response for all interlayer excitons, the maxima of the transmittance peak on both the ports should lie close to the emission wavelength of the dipole. It can be achieved by tuning the coupling gap between the microring resonator and the bus-waveguide. Figure~\ref{fig:figure3}(a) displays the fine tuning of the resonant transmission peak for the $K$ valley exciton at RSP and LSP ports by varying the coupling gap from 200~nm to 280~nm in steps of 40~nm. This can be attributed to coupling induced frequency shift\cite{Li2009}. The resonant peak for the 280~nm gap was found to be the closest to the targeted excitonic wavelength with a very small offset of $\approx0.44$~nm and hence the gap was fixed to 280~nm for all subsequent analysis related to the 908.48~nm interlayer exciton. Figure~\ref{fig:figure3}(d) and (b) shows a pronounced difference in the mode intensity collected at the two end ports of the bus waveguide with RSP being stronger than the LSP for $K$ valley exciton. Figures~\ref{fig:figure3}(e and c) present a similar analysis for the conjugate dipole corresponding to $K'$ valley exciton. Consequently, as the valley hosting the interlayer exciton is changed, the mode intensity or the transmittance at the two ports is also flipped yielding a highly directional separation and routing of the valley state.\\

To quantify the valley routing efficiency of the proposed system, the ratio of the area under the transmittance peaks (at the targetted excitonic emission wavelength) collected at the two ports was calculated. This ratio, from hereon referred to as ``$\rm t_{r}$", gives a measure of the efficiency of valley state routing. $\rm t_{r}$ of $\approx 3.66$ and $\approx 3.06$ was obtained for the $K$ and $K'$ valley excitons, respectively. Moreover, in accordance with the resonant condition for a cavity given by $\rm m\lambda=n_{e f f}L$ (where $L$ represents total optical path length, $\rm n_{e ff}$ is effective refractive index, and $m$ is the resonant order of the cavity) \cite{bogaerts2012silicon}, the ${\rm Si_3N_4}$ microring resonator supports several resonant WGM peaks. The free spectral range (FSR) of the microcavity was estimated by simulating the transmission spectra over a wavelength span of 180~nm around the central wavelength (908.48~nm in this case) and was found to be $\sim$4.04~nm, in good agreement with the theoretical FSR calculation. Figure~\ref{fig:figure3} shows the simulated transmission spectra for $K$ valley exciton, overlapped with the free-space photoluminescence (PL) for ${\rm MoSe_2/WSe_2}$ vdWH at room temperature as measured by Khelifa et al. \cite{khelifa2020coupling}. It highlights that the room temperature PL envelops several cavity modes opening avenues for realizing cavity enhanced vdWH based light sources including nanolasers using our proposed platform. 

\begin{figure*}
    \centering
    \includegraphics[width=\textwidth]{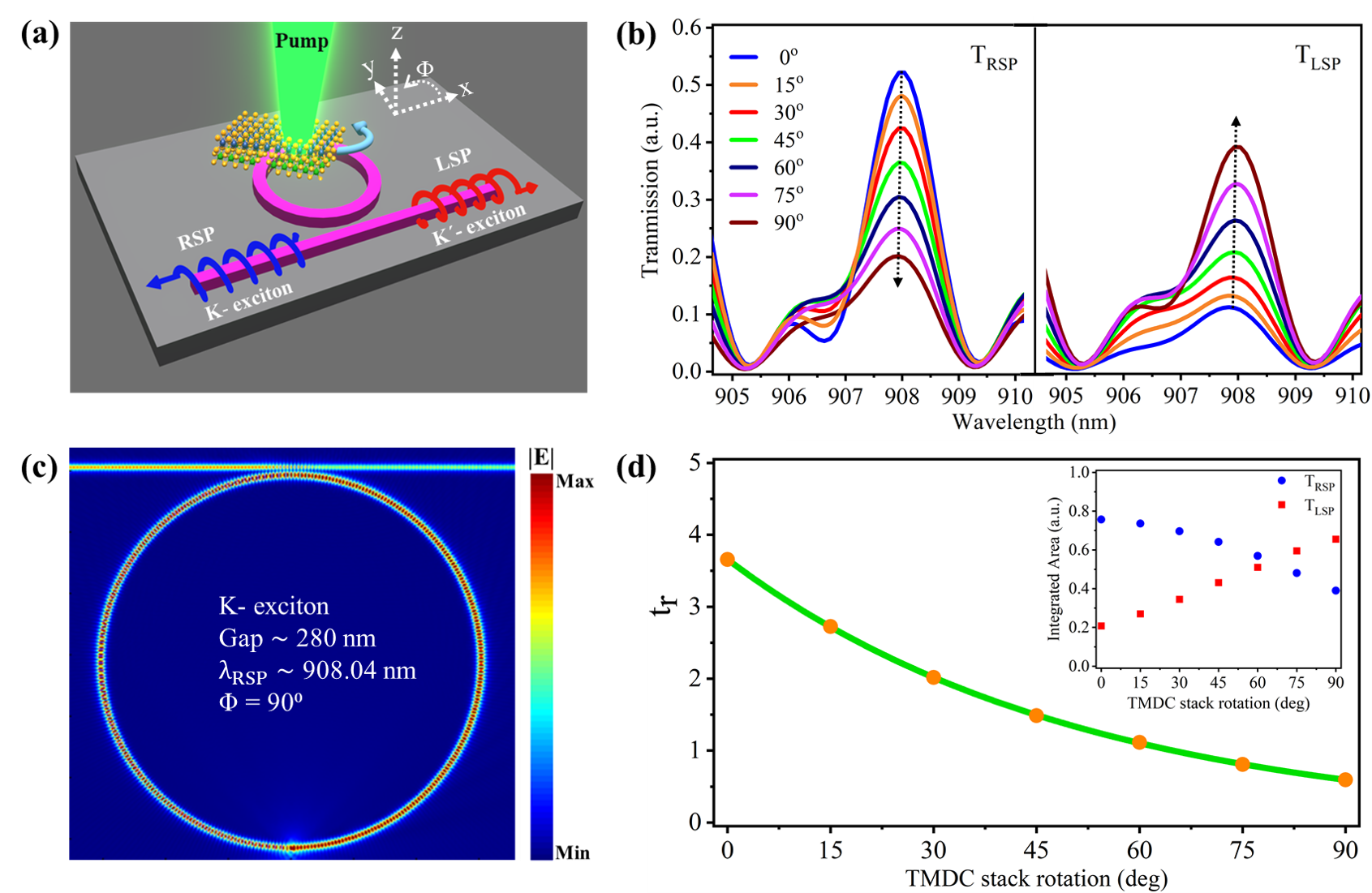}
    \caption{\textbf{Effect of interlayer exciton orientation on valley routing:} Effect of vdWH orientation on the asymmetric transmission response of the coupled K valley interlayer exciton. (a) Illustration of detecting cavity-coupled interlayer exciton by rotating the TMDC stack in xy-plane by angle $\Phi$. (b) Shows the transmission spectra collected at RSP (left side) and LSP (right side) of waveguide at different values of $\Phi$.(c) Depicts the simulated electric field profile, $ \rm |E|$ of the propagating modes for $\Phi$ = 90$^{\circ}$.(d) Shows change in $\rm t_r$ as a function of $\Phi$ with inset showing the variation in relative amplitudes of $\rm T_{RSP}$ and $\rm T_{LSP}$.
    }
    \label{fig:figure4}
\end{figure*}

\emph{Effect of TMDC stack alignment on valley routing--}
So far in this work, the TMDC stack orientation is considered to be aligned with the bus waveguide. However, in actual experiments, this condition might vary and affect the valley state routing efficiency of the device. Previously, Sigl et al.~\cite{sigl2022optical} have shown the effect of optical dipole orientation on the far-field emission of both H-type and R-type stacking configuration for $\text{MoSe}_{2}/\text{WSe}_{2}$ vdWH at low-temperature. Therefore, to study the affect of orientation of TMDC stack in our system, we rotate the TMDC stack in $x-y$ plane, as shown in Figure~\ref{fig:figure4}(a). It is found that as the misalignment between the TMDC stack and x--axis (which coincides with the longer axis of the bus waveguide) is increased by rotating the stack from ${\rm \Phi = 0^\circ}$ to ${\rm \Phi = 90^\circ}$, both forward and backward travelling modes get affected (see Figure~\ref{fig:figure4} (b) and (d)), changing the transmission at the LSP and RSP ports of the bus waveguide. Consequently, the $\rm t_{r}$ values drop significantly from 3.66 to 0.60 as shown in Figure~\ref{fig:figure4}(d). Additionally, we observe a complete flip in transmission at RSP and LSP ports (inset of Figure~\ref{fig:figure4}(d)), which is also evident from electric field ($ \rm |E|$) distribution shown in Figure~\ref{fig:figure4}(c). This suggests that our proposed system would also act as an efficient metrology tool to characterize the orientation of the vdWH stack on the ring waveguide.\\

\emph{Conclusion--}
In this paper, we propose a novel nanophotonic-hybrid device comprising interlayer excitons in $\text{MoSe}_{2}/\text{WSe}_{2}$ vdWH TMDCs stack, coupled with Si$_{3}$N$_{4}$-microring resonator based PIC platform for efficiently routing the valley state. The heterobilayer TMDC stack can be monolithically integrated on the top of Si$_{3}$N$_{4}$-microring resonator at the symmetric position, which is coupled with bus waveguide. For efficient coupling, the waveguide was optimized to operate in single-mode condition. These interlayer excitons ($K$ and $K'$ valley) are circularly polarized with slighlty tilted quantization axes resulting in a highly directional and asymmetric transimission at the waveguide end ports for different valley states of the interlayer excitons. This valley routing phenomenon, quantified using $\rm t_r$, is also found to be strongly dependent on the orientation of the TMDC stack with respect to the bus-waveguide axis -- a property which can enable our proposed system to act as a metrology tool for determining the orientation of the vdWH stack. This approach helps in valley selective routing of interlayer excitons, offering a CMOS compatible platform for future on-chip optoelectronic devices\cite{Lin2021,Mak2018,Wilson2021}.

\emph{Acknowledgment--} A.K. acknowledges funding support from the Department of Science
and Technology via the grants: SB/S2/RJN-110/2017,
ECR/2018/001485 and DST/NM/NS-2018/49.

\bibliography{main-ref,intro}

\end{document}